\documentclass[11pt,twoside]{article}
\usepackage{asp2010}

\resetcounters

\begin{document}

\title{Looking for Systematic Variations in the Stellar Initial Mass Function}
\author{N. Bastian$^1$, K.R. Covey$^2$, and M.R. Meyer$^3$
\affil{$^1$School of Physics, University of Exeter, Stocker Road, Exeter EX4 4QL, UK}
\affil{$^2$Department of Astronomy, Cornell University, Ithaca, NY 14853, USA}
\affil{$^3$Institute of Astronomy, ETH Z\"{u}rich, Wolfgang-Pauli-Str. 27, 8093 Z\"{u}rich, Switzerland}}

\begin{abstract}
Few topics in astronomy initiate such vigorous discussion as whether or not the initial mass function (IMF) of stars is universal, or instead sensitive to the initial conditions of star formation.  The distinction is of critical importance: the IMF influences most of the observable properties of stellar populations and galaxies, and detecting variations in the IMF could provide deep insights into the process by which stars form.  In this contribution, we take a critical look at the case for IMF variations, with a view towards whether other explanations are sufficient given the evidence.  Studies of the field, local young clusters and associations, and old globular clusters suggest that the vast majority were drawn from a ``universal" IMF.  Observations of resolved stellar populations and the integrated properties of most galaxies are also consistent with a ``universal IMF", suggesting no gross variations in the IMF over much of cosmic time. Here we focus on 1) nearby star-forming regions, where individual stars can be resolved to give a complete view of the IMF, 2) star-burst environments, in particular super-star clusters which are some of the most extreme objects in the universe and 3) nearby stellar systems (e.g. globular clusters and dwarf spheroidal galaxies) that formed at high redshift and can be studied in extreme detail (i.e. near-field cosmology).
\end{abstract}

\section{Introduction}


In this contribution, we briefly summarise our recent review on
stellar IMF variations \citep*{bastian10}. The IMF of the
field has been studied extensively (e.g.~\citet{covey08,bochanski10}), and we refer the reader to the contribution
by Bochanski elsewhere in this volume for recent updates.  In this
contribution, we review the evidence for IMF variations in three
distinct Galactic and extra-galactic environments.  In \S~2, we
consider nearby star-forming regions and stellar clusters, where
individual stars can be resolved and the IMF studied in exquisite
detail. In \S~3, we review measurements of the IMF in starburst
galaxies, with particular attention to the super-star clusters they
host, and which represent the most extreme environments in the
Universe with respect to stellar densities and star-formation rate
densities. In \S~4 we consider IMF measurements in the context of
``near-field cosmology", namely using local objects, such as globular
clusters and dwarf spheroidal galaxies, to study the shape of the IMF
at high redshift.

\section{Nearby Star-Forming Regions and Open Clusters}

Nearby open clusters and young star forming regions provide several
advantages for IMF determinations: their proximity allow the IMF to be
probed to the lowest masses via direct star counts; brown dwarfs are
warm and bright at the youngest ages, easing their observational
detection; their physical coherence provide a discrete sampling of the
IMF as a function of metallicity and stellar density; and the short
duration of their star formation events ameliorate ambiguities due to
different potential star formation histories.
For these reasons, open clusters and star forming regions have been
frequent targets for local IMF measurements.  It is beyond the scope
of this contribution to review all such determinations, but an
assortment of recent mass function measurements for star-forming
associations, open and globular clusters is shown in Fig. 1 (from \citealt{demarchi10}). Here, we can see the characteristic shape of the
IMF; a power-law distribution on the high mass side ($>$ 1M$_{\odot}$)
and a lognormal (or multiple power laws) form below with a
characteristic mass of  $\sim0.2 - 0.3~$M$_{\odot}$ below this. The vast
majority of star-forming regions and open clusters appear to have
stellar mass distributions that are qualitatively and quantitatively
consistent with a Kroupa/Chabrier type IMF \footnote{\citet{dab08} have shown that the log-normal parameterization
of the IMF advanced by \citet{chabrier03} is extremely similar to the
two-part power-law defined by \citet{kroupa93}; we
therefore refer to all IMFs of this form as Kroupa/Chabrier-type
IMFs.}.

Most of the IMF variations reported in young regions can be ascribed
to sampling effects, given the finite numbers of stars in each region,
or to systematic effects, arising from differences in the analysis,
such as the prescription adopted to correct for extinction or convert
observational luminosities into model dependent stellar masses.
Hence, it is imperative to systematically compare observations of
different regions before claims of variations are made. 
Variations should be identified not solely on the basis of a
comparison of the IMF index measured in two studies, but rather from a
direct comparison of the underlying observational samples (i.e.
color/luminosity distributions) and/or by analyzing the (model
dependent) mass distribution in a more statistically robust way (i.e.
through KS tests).

A handful of regions have been studied systematically in this way: of
these, the prime example of a deviant IMF is that of the Taurus
star-forming region. Comparing the IMF of Taurus, to those measured
from the ONC, IC 348 and Chameleon I using the same methods, \citet{luhman09} find that Taurus has a significant excess of stars with
mass $\sim$0.7 M$_{\odot}$ compared to these other regions.
Definitively identifying the root cause of this variation (e.g., a
higher Jeans mass), and eliminating the possibility that this is a
rare, statistical fluke, requires the discovery of other regions with
similar physical properties and deviant IMFs, which we currently lack.
To some extent, it is remarkable how well the low-mass end of the mass
function agrees across different star-forming regions. Evolutionary
tracks of low-mass PMS are highly uncertain, making the conversion
from observed color/luminosity to mass (especially without
spectroscopic typing) fraught with difficultly. Stellar rotation,
accretion history and magnetic activity can all influence a star's
position in the CMD, presenting serious, potentially insurmountable,
challenges for using a (pre-main sequence) star's location in the CMD to accurately infer
its age and/or mass (e.g. \citealt{jeffries07,mayne08,baraffe09}). This re-iterates the need for authors to publish full
observational catalogues for their mass function studies, enabling
future tests for mass function variations to be based upon direct
observable properties, rather than on second order inferred
properties.

\begin{figure}[!ht]
\plotone{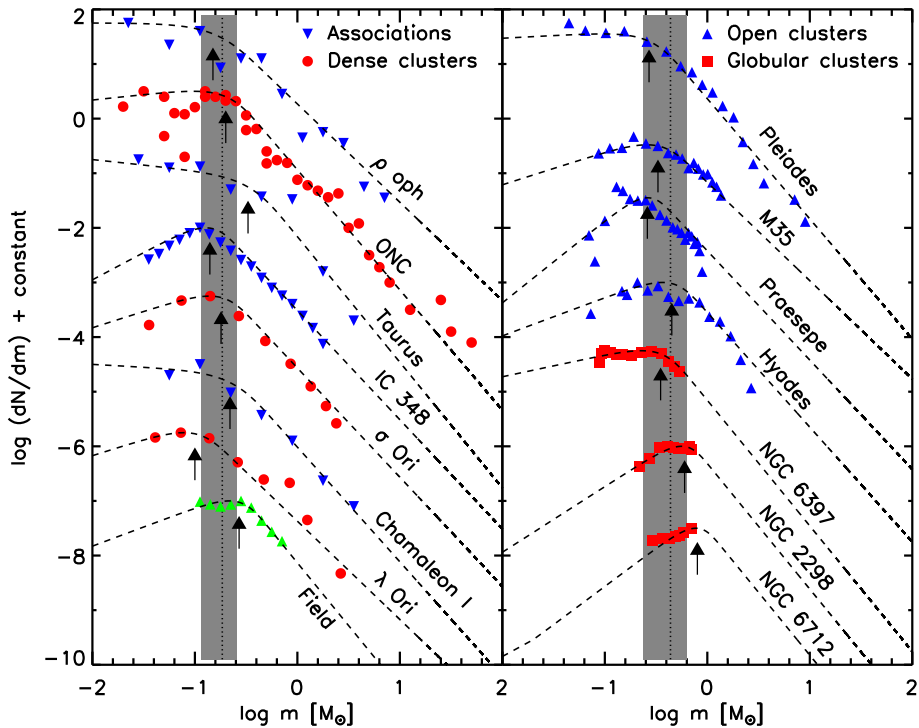}
\caption{The derived present day mass function of a sample of young star-forming regions, open clusters spanning a large age range, and old globular clusters from the compilation of \citet{demarchi10}.  Additionally, we show the inferred field star IMF.  The dashed lines represent ``tapered power-law" fits to the data. The arrows show the characteristic mass of each fit ($m_p$), the dotted line indicates the mean characteristic mass of the clusters in each panel, and the shaded region shows the standard deviation of the characteristic masses in that panel (the field star IMF is not included in the calculation of the mean/standard deviation).  The observations are consistent with a single underlying IMF, although the scatter at and below the stellar/sub-stellar boundary clearly calls for further study.  The shift of the globular clusters characteristic mass to higher masses is expected from considerations of dynamical evolution. See \citet*{bastian10} for details.}
\end{figure}

\section{Starburst Galaxies and Their Super-Star Clusters}

\subsection{The technique and application to SSCs}
It is often stated that the IMF in starburst galaxies is significantly different than that seen locally.  In particular, that the conditions prevailing in starbursts inhibit the formation of low-mass stars, resulting in a ``top-heavy" IMF.  It is difficult to place direct constraints on the IMF within starbursts due to the high star-formation rates, complicated extinction patterns (i.e. dust lanes) and possible presence of an AGN (not to mention their large distances making it impossible to resolve individual stars).  One promising route, however, is through dynamical methods.  This entails measuring the dynamical mass of the system, $M_{\rm dyn}$ (through Jeans modelling or simply adopting Virial equilibrium) and comparing that to what is expected from modelling of the stellar population, $M_{\rm pop}$.  The latter depends on the age and metallicity of the stellar system and crucially on the adopted IMF.  This technique relies on the fact that for ``standard" IMFs the mass of the system is dominated by low-mass stars.  This is demonstrated in Fig.~2.  The light of these systems, however, is dominated by high-mass stars, so in essence this technique compares the mass-to-light ratio of a system to that expected for various IMFs.  

\begin{figure}[!ht]
\plotone{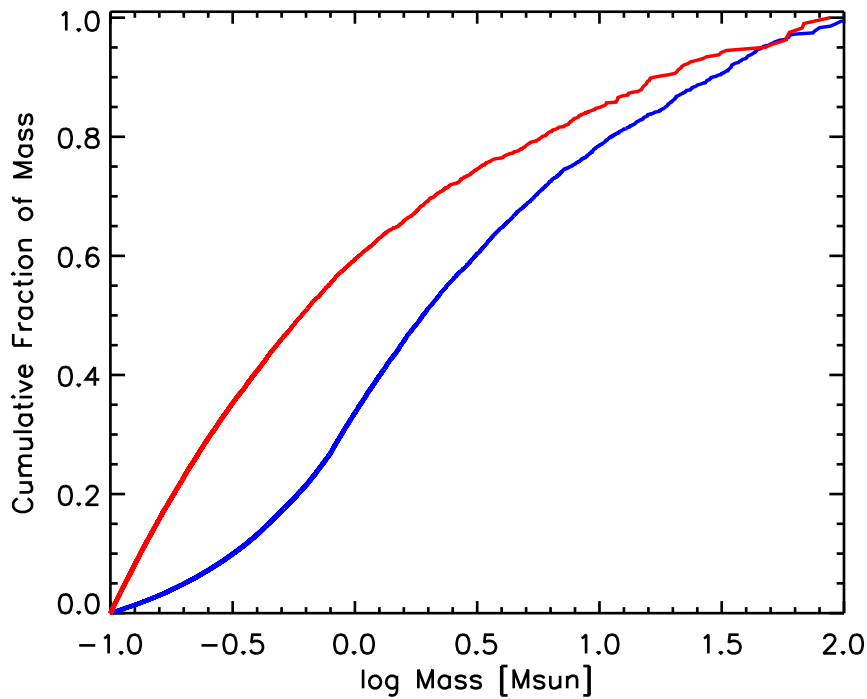}
\caption{The cumulative fraction of mass in a stellar population with a \citet{salpeter55} (red/top line) and \citet{kroupa02} (blue/bottom line) type IMFs.  Note that 50\% of the mass of a (single age) stellar population is made up of stars less than $0.6~M_{\odot}$ in the case of a Salpeter distribution or less than $2.0~M_{\odot}$ in the case of a Kroupa type IMF.  In either case, low mass stars dominate the mass of the stellar system (we have restricted the range considered from 0.1 to 100~M$_\odot$).
}
\end{figure}

The simplest case is for stellar clusters, for which all the stars have the same age and metallicity. Within starburst (and quiescent) galaxies, there exist stellar clusters with high masses (in some cases up to $\sim10^8 M_{\odot}$ - \citealt{maraston04}) and ages between a $<10$ to 1000s of Myr (see \citealt{pz10} for a recent review).  These super-star clusters (SSCs) are thought to be young globular clusters, and their brightness allows their velocity dispersions to be obtained with current instrumentation.  In Fig.~3, a selection of SSCs with estimated dynamical and population masses are shown as solid (red) squares (taken from \citealt{bastian06}).  In the bottom panel, the horizontal lines show the expected ratio of $M_{\rm dyn}/M_{\rm pop}$ if the underlying stellar IMF has the Kroupa/Chabrier (dashed) or the Salpeter (dashed-dotted) form.  Note that the points are consistent with a Kroupa/Chabrier form of the IMF, i.e. the same as that seen locally.  While subtle variations are not ruled out by this method, drastic changes are clearly excluded.

SSCs are a particularly important benchmark to test for IMF variations due to their extreme nature.  Their star-formation rates as well as their stellar surface and volume densities are orders of magnitudes higher than any galaxies in the nearby or distant universe.  The fact that the IMFs of SSCs are similar to that seen in local diffuse star-forming regions argues strongly that the IMF does not change significantly as a function of the local environment.  Hence, it would be strange indeed if trends were found linking the global environment (e.g. galaxy type or galaxy-wide star-formation rate) to IMF variations.

\subsection{Early type galaxies}
In a similar way to cluster investigations, one can carry out the same analysis for entire galaxies.  This is complicated by the fact that galaxies are made up of multiple generations of stars (with different ages/metallicities) and extinction is often non-uniform.  One way to circumvent this problem to some extent is to focus on Early Type Galaxies (ETGs) which are dominated by a single (old) stellar population.  This has been done by e.g. \citet{cap06,cap09}, who studied ETGs at low and high redshift (shown as empty circles and filled triangles respectively in Fig. 3).  The authors conclude that both samples are consistent with a Chabrier/Kroupa type IMF and that there is no deficiency of low mass stars in these (presumably post-starburst) galaxies.

\subsection{Direct detection of low-mass stars in integrated spectra}
With the advent of sensitive near-IR spectrographs on 8-10m class telescopes, it is now possible to detect the spectroscopic features of low-mass pre-main sequence (PMS) stars.  The reason for this is that PMS are orders of magnitude brighter than their main sequence counterparts allowing for the detection of specific spectral features in the integrated light of young stellar clusters \citep{meyer05}\footnote{A similar technique has recently been suggested by \citet{dokkum10}, to investigate the presence of old sub-solar mass stars in elliptical galaxies.}.  One can apply this technique directly to ongoing starbursts (and the stellar clusters within them) in order to detect the presence/absence of low-mass stars, hence directly constraining the IMF.  A first test of this technique has been carried out on young clusters in the Antennae galaxies and low-mass PMS stars were unambiguously detected \citep{greissl10}.  However, due to the relatively large slit-width ($\sim90$~pc) the spectra were contaminated by surrounding (unassociated) RGB stars, hence it was not possible to quantify the numbers of low-mass stars present and constrain the slope of the IMF.  Future studies, concentrating on nearby systems, are a promising way to detect and quantify IMF variations in the low-mass regime in different extragalactic environments.

\begin{figure}[!ht]
\plotone{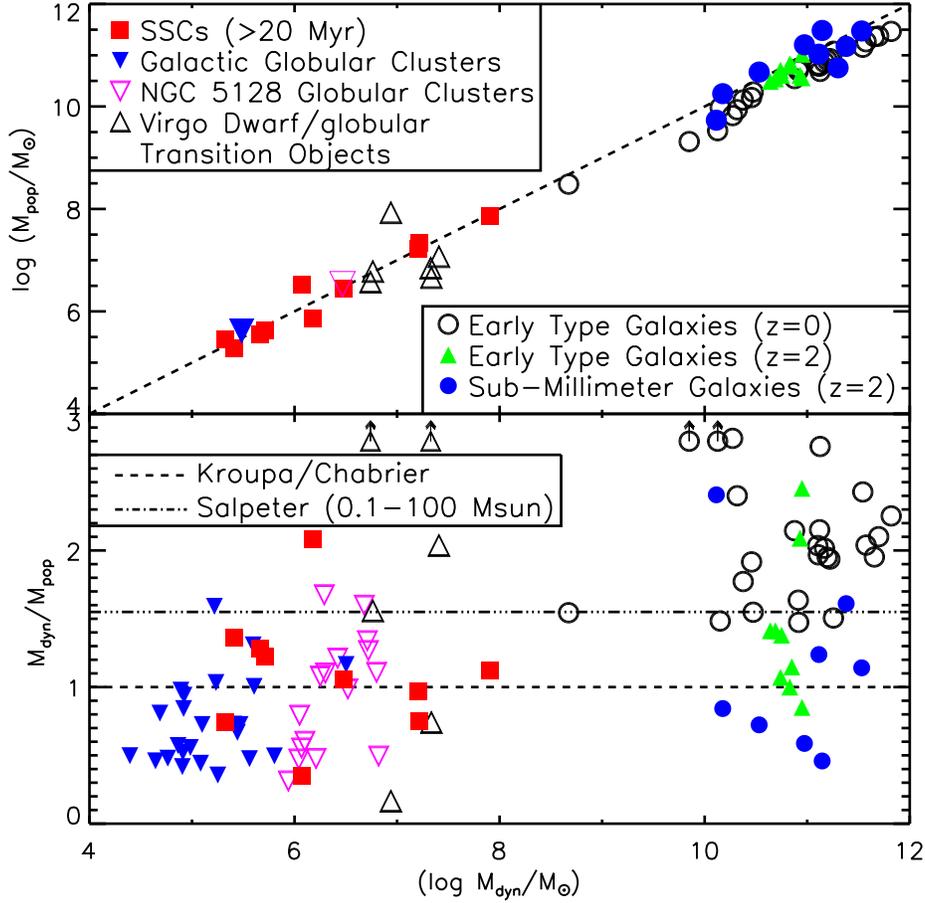}
\caption{{{\bf Top panel:} The measured dynamical mass, $M_{\rm dyn}$ (or through Jeans dynamical modelling in the case of the Early Type Galaxies - see \citealt{cap06}) versus the stellar mass derived through modelling of their integrated light using simple/composite stellar population models which adopt an Kroupa type IMF.  The solid red squares are super star clusters with ages greater than 20~Myr, the upside down blue filled triangle and open magenta upside down triangle represents the mean of 24 Galactic globular clusters and 16 globular clusters in NGC~5128, respectively, open circles show Early Type Galaxies in the local Universe and filled green triangles and blue circles show Early Type Galaxies and Sub-millimeter galaxies at high redshift. {\bf Bottom panel:} The same as the top panel except now the ratio between $M_{\rm dyn}$ and $M_{\rm pop}$ is shown.  If the underlying IMF was well described by a Kroupa-type distribution a ratio of 1 is expected in this representation (shown as a dashed line).  If a Salpeter IMF (down to 0.1~$M_{\odot}$) is a good representation of the underlying IMF a value of 1.55 is expected (dash-dotted line).  Note that the galaxy points are upper limits, as a fraction of the $M_{\rm dyn}$ measurement is expected to come from dark matter.  See \citet*{bastian10} for details.
}}
\end{figure}

\section{Near-Field Cosmology}

While the most massive stars dominate the light that we see in star-forming galaxies at high redshift, it is the low mass stars which will remain for many Gyr, bearing the imprint of their IMF.  Locally, there are abundant examples of the remnants of star formation from the early Universe ($z \ge 3$), namely globular clusters (GCs) and dwarf spheroidal galaxies. 

\subsection{Globular clusters}
Globular clusters (GCs) are thought to be, by and large, made up of a single population of stars with a common age and metallicity.  GCs are some of the oldest objects in the Universe \citep{brodie06}, although GC-like objects continue to form in the present day (see \S~3.1).  Due to the proximity of Galactic GCs, many studies of their mass functions have been carried out using deep HST and ground based imaging.  Fitting their  mass functions leads to characteristic masses (adopting the \citet{chabrier03} form) of $M_{\rm c} =0.33$~\citep{paresce00}.  This
characteristic mass is similar to, although slightly larger than, that
found for young clusters/star-forming regins in the Milky Way disk ($M_{\rm c}=0.1-0.3$, see \S~2).  However, as pointed out by \citet{dokkum08}, this characteristic mass is clearly lower than predicted by the results of \citet{dokkum08} and \citet{dave08} (based on inferred IMF variations using galaxy scaling relations), which expect $M_{\rm c}>4 M_{\odot}$ for the globular cluster formation epoch ($z=3-5$).  Additionally, intermediate age ($4.5$~Gyr, $z_{\rm form} \sim 0.4$) clusters in the SMC also appear to have mass functions consistent with globular clusters and young clusters \citep{rochau07}.

The similarity between the IMF of young clusters and globular clusters is shown in Fig.~1.  Fitting tapered power-laws to the young and old clusters results in a consistent picture where the stars in both young and old clusters formed from the same underlying IMF \citep{demarchi10}.  The characteristic mass in older clusters does appear to be systematically larger than in young clusters and in the field.  However, this is expected from a Hubble time of dynamical evolution, which systematically removes the lower mass members of clusters (e.g. \citealt{baumgardt08,kruijssen09}).

\subsection{Dwarf Spheroidal Galaixes}
Dwarf spheroidal galaxies provide excellent laboratories in which to study the IMF while avoiding the dynamical evolution encountered by GCs. These galaxies have similar ages and star formation histories as GCs, however their stellar densities are orders of magnitude lower, so they experience correspondingly less dynamical evolution.  \citet{wyse02} used HST data to construct a luminosity function (LF) of stars in Ursa Major, an old, local group dwarf spheroidal galaxy.  To avoid the complication of transforming observed luminosities to stellar mass, the authors directly compare the Ursa Major LF to that of globular clusters of similar metallicity.  They find that the LFs are indistinguishable down to $\sim 0.3 M_{\odot}$ , and conclude that the IMF must be independent of density and the presence of dark matter into at least the sub-solar mass regime.

Of course the downside to such investigations is that we are limited to the low-mass regime, as higher mass stars have already ended their lives.  However, the similarity in the low-mass regime and the characteristic mass already places severe restrictions on theories of IMF variations as a function of redshift, metallicity, stellar density, and the presence/absence of dark matter.

\section{Future Work}

Upcoming advances in ground/space based instrumentation as well as theoretical and computational power will lead to significantly tighter constraints on the form of the IMF and any potential variations as a function of environment that may exist.  How exactly is IMF sampled?  What is responsible for setting detailed shape?  Are regions like Taurus truly different, or merely statistical flukes?  Can ``extreme"  local environments (e.g. the Galactic center - \citealt{bartko10}) be used to test/rule-out competing star-formation theories?  These are just some of the questions that will be addressed in detail over the next few years.

Realizing the promise of these technical advances, however, will require a similar advance in the statistical analysis of IMF measurements. As we enter this new era, we advocate a shift in the means used to characterize and search for variations in the IMFs of resolved stellar populations. Specifically, we recommend that future IMF studies publish their derived space densities, such that IMF variations can be tested by using a direct statistical comparison of two measured IMFs, such as with a KS test, rather than by comparing the parameters of the analytic fit adopted to characterize these increasingly rich datasets. If a functional form is fit to a IMF measurement, we suggest that statistical tools such as the F-test can provide quantitative guidance as to the most appropriate functional form to adopt, and that the uncertainties associated with the derived parameters be clearly reported. By providing a more statistically sound basis for IMF comparisons, we will be better poised to uncover IMF variations where they do exist, and quantify the limits on IMF variations imposed by measurements consistent with a ``universal IMF".

\bibliography{bastian_n}

\end{document}